\documentclass[apjl]{emulateapj}                                                       % FROM papers/26/letter/rv.tex
\shorttitle{GEOMETRY AND HIGH-Z 21-CM ABSORPTION}
\shortauthors{S. J. CURRAN}

\def\kms{km ${\rm s}^{-1}$}

\def\ch2{$\chi^2$}

\def\kms {\hbox{${\rm km\ s}^{-1}$}}

\def\scm  {$\hbox{{\rm cm}}^{-2}$}    %cm-2
\def \HI {H{\sc \,i}}

\def\lapp{\ifmmode\stackrel{<}{_{\sim}}\else$\stackrel{<}{_{\sim}}$\fi}
\def\gapp{\ifmmode\stackrel{>}{_{\sim}}\else$\stackrel{>}{_{\sim}}$\fi}

\begin{document}


%\title{The geometry of a flat expanding Universe and the detection of cool neutral gas at high redshift}
% The effects of geometry on the detection of cool neutral gas at high redshift   ??? 
\title{The geometry effects of an expanding Universe on the detection of cool neutral gas at high redshift}


%\author{S. J. Curran\altaffilmark{1} and M. T. Whiting\altaffilmark{2}}
\author{S. J. Curran}
\affil{Sydney Institute for Astronomy, School of Physics, The University of Sydney, NSW 2006, Australia}
\email{sjc@physics.usyd.edu.au}

%\and
\affil{ARC Centre of Excellence for All-sky Astrophysics (CAASTRO)}

\begin{abstract}
Recent high redshift surveys for 21-cm absorption in damped Lyman-$\alpha$ absorption systems (DLAs)
take the number of published searches at $z_{\rm abs} > 2$ to 25, the same number as at $z_{\rm abs} < 2$,
although the detection rate at high redshift remains significantly lower (20\% cf. 60\%).
Using the known properties of the DLAs to estimate the unknown profile widths of the 21-cm non-detections and including the limits
via a survival analysis, we show that the mean spin temperature/covering factor degeneracy at high redshift is, on average, 
double that of the low redshift sample. This value is significantly lower than the previous factor of eight for the spin temperatures
 and is about the same factor as in the angular diameter distance ratios between
the low and high redshift samples. That is, without
the need for the several pivotal assumptions, which lead to an evolution in the spin temperature, we show that
the observed distribution of 21-cm detections in DLAs can be accounted for by the geometry effects of an expanding Universe.
That is, as yet there is no evidence of the spin temperature of gas rich galaxies evolving with redshift.

\end{abstract}


\section{Introduction}
\label{intro}

The 21-cm transition of hydrogen (\HI) traces the cool component of neutral gas throughout the Universe,
the raw material fuelling star formation.
In emission,  the low
probability of the 21-cm spin-flip transition compounded by the inverse square law, renders this essentially
undetectable at $z\gapp0.2$. In absorption, however, this sensitivity limit can be avoided since the line strength is independent
of distance, depending only upon the column density of the absorbing gas and the flux from the background emission
source. Nevertheless, detections remain rare with only 77 \HI\ 21-cm absorption systems known at $z\geq0.1$ --- 35 {\em
  associated} with the quasar host galaxy ($z_{\rm abs} = z_{\rm QSO}$)\footnote{Summarised in \cite{cw10}, with the
  addition of three new associated absorbers reported in \citet{cwm+10,cwwa11}.},  in addition to 42 systems located at
some intermediate redshift along the sight-line to the quasar ($z_{\rm abs} < z_{\rm QSO}$)\footnote{Summarised in
  \cite{cur09a}, with the addition of those reported by \cite{cwt+11,sgp+12}.}. These {\em intervening} absorbers occur
in known or candidate\footnote{Usually in Mg{\sc \,ii} absorption systems at $z_{\rm abs}\lapp1.7$, thus not being
  sufficiently redshifted to be detected in the optical band by ground-based telescopes.}  damped Lyman-$\alpha$
absorption systems (DLAs, defined by their large neutral hydrogen column densities, $N_{\rm
  HI}\geq2\times10^{20}$ \scm). Currently there are 1500 DLAs known \citep{cwbc01,npls09}, which could
contain over 80\% of the neutral gas mass density of the Universe \citep{phw05}. 

The velocity integrated optical depth of the 21-cm absorption, $\tau\equiv-\ln\left(1-\frac{\sigma}{f\,S}\right)$, is
related to the total neutral hydrogen column density via \citep{wb75}
\begin{equation}
N_{\rm HI}=1.823\times10^{18}\,T_{\rm spin}\int\!\tau\,dv\,,
\label{enew}
\end{equation}
where $T_{\rm spin}$ [K] is the mean harmonic spin temperature of the gas, $\sigma$
is the depth of the line (or r.m.s. noise in the case of a non-detection) and $S$ \& $f$ the flux density and covering factor of
the background continuum source, respectively. This latter parameter accounts for the possibility that the absorbing gas 
may not fully cover all of the background radio emission\footnote{Optical spectra from QSOs come from
  continuum sources $\lapp1$ pc in diameter, whereas radio sources can exceed $\sim100$ pc (see \citealt{wgp05}).}
and so all of the measured flux, $S$, may not be intercepted, decreasing the sensitivity of the observation.

In both the associated and intervening cases the detections occur overwhelmingly at $z\lapp1$. For the associated systems,
$N_{\rm HI}$ is generally not known, although \citet{cww+08} attribute the lack of absorption 
at high redshift to the traditional optical selection of targets biasing searches towards the
most luminous objects at these large distances, where  the ultra-violet luminosity ionises the neutral gas.
 
In the optically thin regime ($\sigma\ll f.S$),\footnote{Which applies to all but one of the known 21-cm absorbers
  \citep{rbb+76}.} Equ. \ref{enew} reduces to $N_{\rm HI}=1.823\times10^{18}\frac{T_{\rm spin}}{f}\int\!\frac
{\sigma}{S}\,dv$, thus giving a degeneracy in the spin temperature/covering factor ratio ($T_{\rm spin}/f$), for the DLAs where the column
density is known. Therefore, 
since there is a mix of detections and non-detections in DLAs at $z_{\rm abs}\lapp2$, whereas at $z_{\rm abs}\gapp2$ \HI\ 21-cm
tends not to be detected, \citet{kc02} suggest an evolution in the spin temperature of the gas, where
this is exclusively high at high redshift. This has direct bearing on the relative proportions of the cold and
warm neutral media (e.g. \citealt{kc02,sgp+12}) and thus the reservoir of raw material available for star formation in 
the early Universe.

However, the conclusion that the spin temperature evolves with redshift
relies on the assumption of a covering factor for each DLA, and
\citet{cmp+03} have shown that, if the degeneracy is left intact, $T_{\rm spin}/f$ is not the order of magnitude
larger at $z_{\rm abs}\gapp2$, as suggested by the $T_{\rm spin}$ distribution of \citet{kc02}. Furthermore, \citet{cw06}
suggest that geometry effects introduced by an expanding Universe could contribute
to the distribution, indicating that the covering factor may actually  be the dominant parameter. Since then
there have been further high redshift searches \citep{ctd+09,sgp+12}\footnote{From the Sloan Digital Sky Survey DLA  Data Releases 1--3 \citep{phw05}  and the SDSS DR7 DLA Data Release 7 \citep{npls09}, respectively.}. This takes the number
of confirmed DLAs and sub-DLAs searched for 21-cm to 25 at $z\geq2$, the same number as at $z\leq2$, thus allowing us to adequately address
the spin temperature versus covering factor  issue.

\section{New \HI\ 21-cm absorption search results}
\label{new}

Although
there are now equal numbers of 21-cm searches above and
below $z_{\rm abs} = 2$, the detection rate at $z_{\rm abs} > 2$ remains considerably lower (5 out of 25) than
for the lower redshift sample (14 out of 25). Since the velocity integrated optical depth of the 21-cm absorption,
normalised by the total neutral hydrogen column density, is proportional to $f/T_{\rm spin}$ (Sect. \ref{intro}), this
suggests that either $T_{\rm spin}$ is generally higher at $z_{\rm abs} > 2$ or that $f$ is generally lower, or a
combination of both, elevating the ratio at high redshift,

In order to obtain limits to the line strengths for the non-detections we must assign a profile width to each absorber. 
This is usually assumed, but here 
we estimate the full-width half maximum (FWHM) of the profile from the 21-cm FWHM---Mg{\sc \,ii} 2796 \AA\ equivalent
width correlation for the detections, specifically  FWHM [\kms]\,$\approx13\,\times{\rm W}_{\rm r}^{\lambda2796}$ [\AA] \citep{ctp+07}.
Where the Mg{\sc \,ii} equivalent widths are not available (generally at $z_{\rm abs}\gapp2.2$ where the transition
is redshifted out of the optical band), we estimate the putative FWHM via the metallicity---Mg{\sc \,ii} equivalent
width correlation \citep{mcw+07}, specifically ${\rm W}_{\rm r}^{\lambda2796}\approx2.0\,{\rm [M/H]} + 4.0$ 
\citep{ctp+07}. When neither the Mg{\sc \,ii} equivalent width nor metallicity are available, we assign a FWHM
of $25$ \kms, the mean of the detections.\footnote{Excluding the FWHM = 240 \kms\ outlier due to the absorber
in the complex sight-line towards 3C\,336 \citep{ctm+07}.}

Given the FWHM of the putative line, in order to obtain the best signal-to-noise ratio, based upon the detection of
a single channel we would smooth the data to a spectral resolution of $\Delta v=\,$FWHM. 
 This has the effect of
improving/degrading the measured r.m.s noise, and thus the optical depth, by a factor of $\sqrt{\Delta v/{\rm FWHM}}$.
The newly derived  integrated optical depth of the line is thus 
\begin{eqnarray}
\left[\int\tau dv \right]_{\rm new}\approx \tau_{\rm peak}\times \sqrt{\frac{\Delta v}{\rm FWHM}}  \times {\rm FWHM} \nonumber \\ = \tau_{\rm peak}\times \sqrt{\Delta v\,{\rm FWHM}},
\end{eqnarray}
where $\tau_{\rm peak}$ is the peak optical depth of the putative line. This has the overall effect of scaling the previously published 
integrated optical depths, where these are calculated per channel,  by $\left[\int\tau dv \right]_{\rm new}/\left[\int\tau dv \right]_{\rm previous}$
\begin{eqnarray}
=\frac{\tau_{\rm peak}\times \sqrt{\Delta v\,{\rm FWHM}}}{\tau_{\rm peak}\times \Delta v} = \sqrt{\frac{\rm FWHM}{\Delta v}}.
\end{eqnarray}
\begin{figure}
\centering \includegraphics[angle=270,scale=0.45]{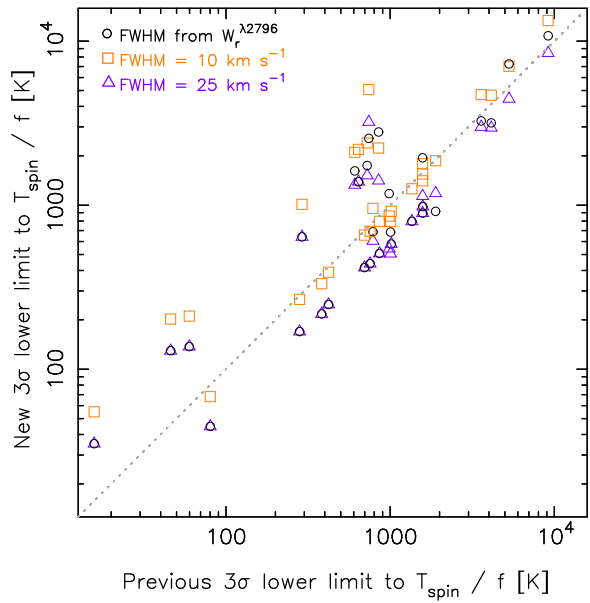}
\caption{The lower limits to the spin temperature/covering factor degeneracy obtained from the 
recalculated velocity integrated optical depths versus those obtained from the previously published values.
The circles show the values obtained by estimating the FWHM from the Mg{\sc \,ii} equivalent width, the
squares using a FWHM of 10 \kms\ and the triangles using the mean 25 \kms\ of
the detections. This is the default FWHM used when the Mg{\sc \,ii} equivalent width or metallicity is
not available, thus giving the overlap between circles and triangles. The dotted line shows where the new and previous values are equal.}
\label{comp-spin}
\end{figure}
We therefore recalculate the limits from the published r.m.s. noise levels and spectral resolutions and in
Fig. \ref{comp-spin} show the effect this has on the $T_{\rm spin}/f$ limits compared to the previous values. For the
method prescribed above (shown as circles), the largest change is for the $z_{\rm abs} = 0.5579$ absorber towards
PKS\,0118--272, for which the limit climbs from $T_{\rm spin}/f > 850$~K \citep{kc01a} to $>1900$ K and, on average, the scaled
values are 160~K ``warmer'' than those published.  As seen from the figure, however, the new limits are mostly
concentrated close the $\left[\int\tau dv \right]_{\rm new} = \left[\int\tau dv \right]_{\rm previous}$ line, with equal
numbers of points to either side.

\citet{ctp+07} attribute the correlation between the velocity spreads of the 21-cm and Mg{\sc \,ii} profiles to the
neutral and singly ionised gas being spatially coincident in 21-cm detected DLAs.
However, given the possibility that much of the ionised
magnesium could arise in warm photoionised gas, and is thus not tracing the same sight-line as the 21-cm, in
Fig. \ref{comp-spin} we also show the $T_{\rm spin}/f$ limits derived from assuming a single FWHM.  For 10 \kms\
(e.g. \citealt{sgp+12}), we find the largest difference to be 4300~K, with a mean rise of 600~K and for 25 \kms, the
largest difference is 2500~K, with a mean decrease of 120~K. Thus, scalings implementing the mean FWHM of the detections have
less of an effect than using 10 \kms, although
either method is preferable to the inhomogeneously calculated published values, since each is consistent in smoothing the data to the FWHM and thus calculating the limit per smoothed channel.

\begin{figure*}
\centering \includegraphics[angle=270,scale=0.70]{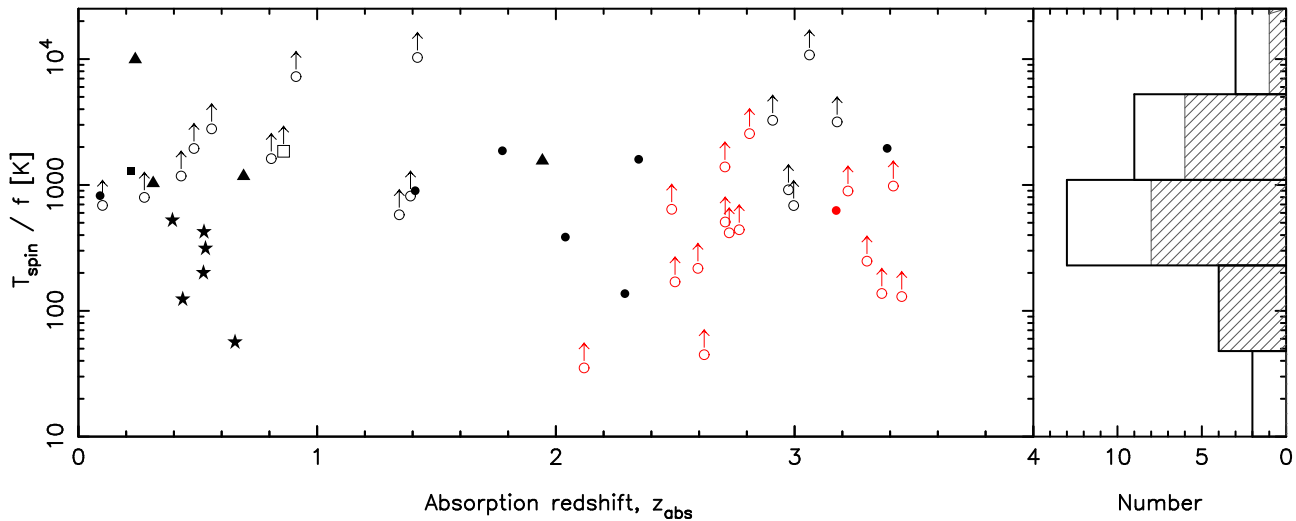}
\caption{The spin temperature/covering factor degeneracy 
versus the absorption redshift for DLAs searched in 21-cm absorption. The filled symbols/histogram represent the 
detections, the unfilled symbols/histogram the non-detections ($3\sigma$ limits),
with the shapes representing the type of galaxy with which
  the DLA is associated: circle--unknown type, star--spiral,
  square--dwarf, triangle--low surface brightness (LSB). The coloured
symbols show the recent high redshift searches of SDSS DLAs \citep{ctd+09,sgp+12}.  In the histogram the limits to $T_{\rm spin}/f$ are treated as actual values.}
\label{Toverf}
\end{figure*}

In Fig. \ref{Toverf} we show the newly derived limits against the absorption redshift.
Applying  a survival analysis to the $3\sigma$ lower limits in $T_{\rm spin}/f$, via the {\sc asurv} package \citep{ifn86}, we obtain mean values of
$\overline{\log_{10} (T_{\rm spin}/f)}_{z\,<\,2} = 3.25\pm0.15$ and $\overline{\log_{10} (T_{\rm spin}/f)}_{z\,>\,2} = 3.56\pm0.17$. 
That is, the mean value 
of $T_{\rm spin}/f$ at $z_{\rm abs} > 2$  is double that at $z_{\rm abs} <2$ (3600 cf. 1800 K).  This is considerably less than the ratio
of 8.0 between the two redshift regimes (8100 cf.  1020 K) found by applying the survival analysis to the spin temperatures of \citet{kc02}.
Since our recalculation of the limits has actually slightly increased the mean value of the spin temperature/covering factor degeneracy, we attribute this
significant difference to the covering factor being left as a free parameter.

\section{Spin Temperature or Covering Factor}

The addition of the new data, together with our treatment of the non-detections, indicates that a factor
of only $\approx2$
in the spin temperature/covering factor ratio between the two
redshift regimes need be accounted for. However, it remains unclear whether $T_{\rm spin}$ or $f$ is the dominant factor.

There have been several attempts to quantify the covering factor through Very Long Baseline Array  (VLBA) imaging
 of the background quasar, in which $f$ is estimated as the ratio of the compact unresolved component's
flux to the total radio flux (e.g. \citealt{bw83,klm+09,sgp+12}). From this,
\citet{klm+09} conclude that for the 21-cm searches, $f\geq0.4$ and hence
low covering factors cannot explain the dearth of detections at high redshift.
However, this method gives no information on the extent of the
absorber nor on how effectively it covers the emission.
Neither does it give information on the depth of the line when the extended
continuum emission is resolved out.

\begin{figure}%[h]
\vspace{3.0cm}
\includegraphics{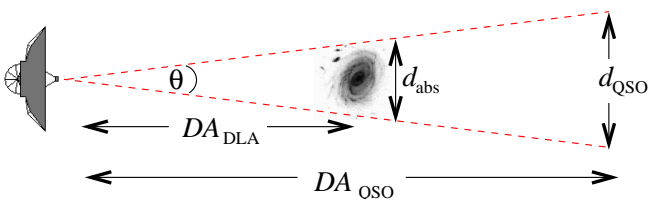} % dla_schem.eps
\caption{Absorber and quasar cross-sections with respect to the angular diameter distances. If the angle required
to fully subtend $d_{\rm abs}$ is less than that to subtend $d_{\rm QSO}$ then $f<1$.}
\label{dla_schem}
\end{figure}

So although an evolution in spin temperature with redshift is implied, this method invokes several crucial assumptions
which could even lead to unphysical results (e.g. the $\sim0$ pc obtained for the extent of PKS 0405--331 at 398 MHz,
\citealt{klm+09}).  On the other hand, due to the effects of an expanding Universe, objects at $1\lapp z\lapp4$ are
always close to a single angular diameter distance ($\sim1500$~Mpc), irrespective of redshift
(e.g. \citealt{hog00}). Beyond $z\sim1$, the angular diameter distance {\em decreases} with increasing redshift,
which will lead to a natural imbalance between the low and high-z DLA samples.
In the small angle approximation the covering factor can be defined as 
\begin{equation}
f\equiv\frac{d_{\rm abs}^{~~~2}}{DA_{\rm DLA}^{~~~2}.\theta_{\rm QSO}^{~~~2}} = \left(\frac{d_{\rm abs}\,DA_{\rm QSO}}{d_{\rm QSO}{DA_{\rm DLA}}}\right)^{2} ,
\label{f}
\end{equation}
where $\theta_{\rm QSO}$ is the angle subtended by the background emission and $d_{\rm abs}$ and $d_{\rm QSO}$ the linear sizes of the cool absorbing gas and background 21-cm emission, respectively (Fig.~\ref{dla_schem}).
Since absorbers at
$z\gapp1$ are always at a similar angular diameter distance as the background quasar ($DA_{\rm DLA} \approx DA_{\rm QSO}$),
these will therefore,  for a given system,
always have a lower coverage than a  counterpart with $DA_{\rm DLA} < DA_{\rm QSO}$ \citep{cw06}, an inequality which can only exist at
$z\lapp1$.

\begin{figure*}
\centering \includegraphics[angle=270,scale=0.70]{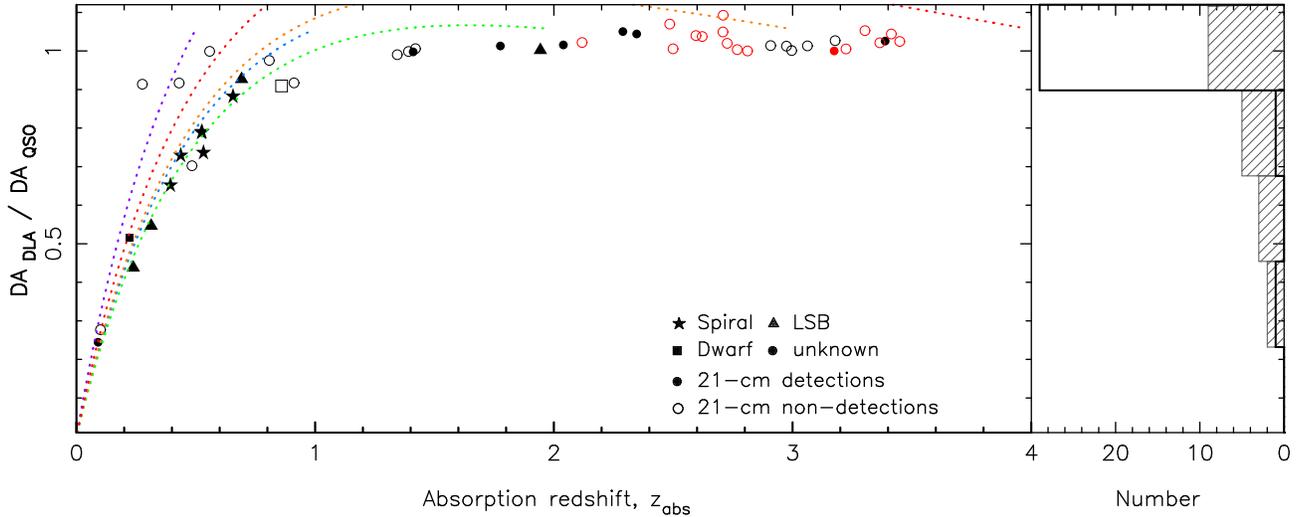} %{distance-z.eps}
\caption{The absorber/quasar angular diameter distance ratio versus the absorption redshift for DLAs searched in 21-cm absorption. 
The symbols and histogram are as per Fig. \ref{Toverf}. The iso-redshift curves show how $DA_{\rm DLA}/DA_{\rm QSO}$ varies with
absorber redshift, where $DA_{\rm QSO}$ is for a given QSO redshift,  given by the terminating value of $z_{\rm abs}$.  That is, we show $DA_{\rm DLA}/DA_{\rm QSO}$ for $z_{\rm QSO}$ = 0.5, 1, 2, 3 and 4. We employ a standard $\Lambda$ cosmology with $H_{0}=71$~km~s$^{-1}$~Mpc$^{-1}$, $\Omega_{\rm
    matter}=0.27$ and $\Omega_{\Lambda}=0.73$. }
\label{distance-z}
\end{figure*}

In Fig. \ref{distance-z} we show the angular diameter distance ratio, $DA_{\rm DLA}/DA_{\rm QSO}$, versus the absorption redshift
for the 21-cm searches.
Note that this distribution is free from any assumptions (including our own with regard to the FWHMs of the non-detections).
From this we see a mix of $DA_{\rm DLA}/DA_{\rm QSO}$ at low redshift (the low values are due to $z\lapp1$ DLAs towards
much higher redshift quasars), whereas values at $z\gapp1$ are exclusively high. That is, a reproduction of the ``spin temperature'' 
distribution of \citet{kc02}. At least the pattern if not the magnitude, given that \citet{kc02} suggest that the average spin
temperature at $z_{\rm abs} > 2$ is eight times that at $z_{\rm abs} < 2$, which is clearly not reproduced in Fig. \ref{distance-z}.

Referring to Sect.~\ref{new}, however, we only require $T_{\rm spin}/f$ at $z_{\rm abs} > 2$ to be about double that
at $z_{\rm abs} < 2$ in order to account for the distribution. From Equ.~\ref{f}, for a given absorber and emitter size,
$f\propto 1/(DA_{\rm DLA}/DA_{\rm QSO})^{2}$, and we find mean values of $\overline{(DA_{\rm DLA}/DA_{\rm
    QSO})^{2}}_{z\,<\,2} =0.68$ and $\overline{(DA_{\rm DLA}/DA_{\rm QSO})^{2}}_{z\,>\,2} =1.06$, a ratio of 1.6. It is
therefore apparent that the distribution in angular diameter distance ratios, and their effect on the covering factor,
is close to that required to account for the  observed $T_{\rm spin}/f$ distribution.

Updating the statistics with the new results, we find a binomial probability of $5.52\times10^{-5}$ of obtaining at
least 9 out of 11 detections at $DA_{\rm abs}/DA_{\rm QSO}<0.8$ and 29 out of 39 non-detections at $DA_{\rm abs}/DA_{\rm
  QSO}>0.8$, by chance.\footnote{$DA_{\rm abs}/DA_{\rm QSO}=0.8$ is the cut used by \citet{cw06} and, although arbitrary,
  this is the lowest ratio which gives an appreciable enough sample size in the lower bin. When a ratio of 0.9 is used
  there are 38 DLAs with $DA_{\rm abs}/DA_{\rm QSO}>0.9$ and 12 with $DA_{\rm abs}/DA_{\rm QSO}<0.9$, which increases
  the significance of the correlation to $4.31\sigma$.} 
That is, a significance of $4.03\sigma$,
lending
weight to the argument that large angular diameter distance ratios are correlated with the non-detection of 21-cm
absorption.  Since there is no physical link between the spin temperature and angular diameter distance ratio, this
suggests that, in the absence of any knowledge of the absorption cross-section in relation to the emitter extent,
the covering factor plays a much more major r\^{o}le than the spin temperature in the detection of 21-cm absorption.

\section{Discussion and Summary}

With the advent of 21-cm absorption searches in high redshift DLAs  there has been much debate
on whether an evolution in spin temperature or lower covering factors are responsible for the 
relatively low number of detections at $z_{\rm abs} > 2$. New searches of SDSS DLAs, which occult the
sight-lines to radio-loud quasars, take the number searched at these redshifts to the same as 
at $z_{\rm abs} < 2$, giving much needed data with which to address this issue.
By giving the 21-cm non-detections a more thorough analysis than in other studies,
through  estimating the FWHMs of the non-detected profiles from the observed
optical properties and applying a survival analysis to the lower limits in $T_{\rm spin}/f$, we find:
\begin{enumerate}
  \item That mean the spin temperature/covering factor ratio at $z_{\rm abs} > 2$ is only double that
found at $z_{\rm abs} < 2$.
This is  significantly lower than the
factor of eight in spin temperature, determined from previous estimates of the covering factor.

\item This reduced factor can be accounted for by the effects of an expanding Universe,
  where the angular diameter distance at $z_{\rm abs} > 1$ is always close to, and sometimes exceeds, that
of the background radio source. That is, the high redshift DLAs are {\em always} disadvantaged in
how effectively they can intercept the background radio flux for a given DLA--QSO pair.

\item Since the non-detections tend to be associated with high ratios in the DLA--QSO angular diameter distances ($DA_{\rm
    abs}/DA_{\rm QSO}\approx1$), whereas the detections are associated with low ratios ($DA_{\rm abs}/DA_{\rm QSO} <
  1$), at {\em all} redshifts, this strongly suggests that the coverage of the background radio source by the absorber
  is the dominant factor.  
\end{enumerate}
This runs contrary to previous studies which estimate the covering factor via VLBA imaging,
  which support an evolution in the spin temperature. However, these estimates are based upon several pivotal
  assumptions, whereas the angular diameter distance ratios are not.

We therefore suggest that there is no evidence for the spin temperature to increase with redshift,
which implies no evolution in the fraction of the warm neutral gas content of DLAs.
\citet{ctp+07}
find a $T_{\rm spin}/f$--metallicity anti-correlation. If the spin temperature does not evolve
with redshift, as does the metallicity \citep{pgw+03,cwmc03},  this may imply that the metallicity is proportional
to the covering factor, which could be due to the larger absorbing cross-sections of the larger galaxies, as evident through the 
spirals which have the largest metallicities and the lower values
of $T_{\rm spin}/f$ \citep{ctd+09}. \\

This research has made use of the NASA/IPAC Extragalactic Database
(NED) which is operated by the Jet Propulsion Laboratory, California
Institute of Technology, under contract with the National Aeronautics
and Space Administration. 
This research has
also made use of NASA's Astrophysics Data System Bibliographic
Service and {\sc asurv} Rev 1.2 \citep{lif92a}, which implements the 
methods presented in \citet{ifn86}.
The Centre for All-sky Astrophysics is an Australian Research Council Centre of Excellence, funded by grant CE11E0090.
%\bibliographystyle{apj}
%\bibliography{aa,ref}

\begin{thebibliography}{28}
\expandafter\ifx\csname natexlab\endcsname\relax\def\natexlab#1{#1}\fi

\bibitem[{Briggs \& Wolfe(1983)}]{bw83}
Briggs, F.~H. \& Wolfe, A.~M. 1983, ApJ, 268, 76

\bibitem[{Curran(2010)}]{cur09a}
Curran, S.~J. 2010, MNRAS, 402, 2657

\bibitem[{Curran {et~al.}(2005)Curran, Murphy, Pihlstr\"{o}m, Webb, \&
  Purcell}]{cmp+03}
Curran, S.~J., Murphy, M.~T., Pihlstr\"{o}m, Y.~M., Webb, J.~K., \& Purcell,
  C.~R. 2005, MNRAS, 356, 1509

\bibitem[{Curran {et~al.}(2010)Curran, Tzanavaris, Darling, Whiting, Webb,
  Bignell, Athreya, \& Murphy}]{ctd+09}
Curran, S.~J., Tzanavaris, P., Darling, J.~K., Whiting, M.~T., Webb, J.~K.,
  Bignell, C., Athreya, R., \& Murphy, M.~T. 2010, MNRAS, 402, 35

\bibitem[{Curran {et~al.}(2007{\natexlab{a}})Curran, Tzanavaris, Murphy, Webb,
  \& Pihlstr\"{o}m}]{ctm+07}
Curran, S.~J., Tzanavaris, P., Murphy, M.~T., Webb, J.~K., \& Pihlstr\"{o}m,
  Y.~M. 2007{\natexlab{a}}, MNRAS, 381, L6

\bibitem[{Curran {et~al.}(2007{\natexlab{b}})Curran, Tzanavaris, Pihlstr\"{o}m,
  \& Webb}]{ctp+07}
Curran, S.~J., Tzanavaris, P., Pihlstr\"{o}m, Y.~M., \& Webb, J.~K.
  2007{\natexlab{b}}, MNRAS, 382, 1331

\bibitem[{Curran \& Webb(2006)}]{cw06}
Curran, S.~J. \& Webb, J.~K. 2006, MNRAS, 371, 356

\bibitem[{Curran {et~al.}(2002)Curran, Webb, Murphy, Bandiera, Corbelli, \&
  Flambaum}]{cwbc01}
Curran, S.~J., Webb, J.~K., Murphy, M.~T., Bandiera, R., Corbelli, E., \&
  Flambaum, V.~V. 2002, PASA, 19, 455

\bibitem[{Curran {et~al.}(2004)Curran, Webb, Murphy, \& Carswell}]{cwmc03}
Curran, S.~J., Webb, J.~K., Murphy, M.~T., \& Carswell, R.~F. 2004, MNRAS, 351,
  L24

\bibitem[{Curran {et~al.}(2011{\natexlab{a}})Curran, Whiting, Murphy, Webb,
  Bignell, Polatidis, Wiklind, Francis, \& Langston}]{cwm+10}
Curran, S.~J., Whiting, M., Murphy, M.~T., Webb, J.~K., Bignell, C., Polatidis,
  A.~G., Wiklind, T., Francis, P., \& Langston, G. 2011{\natexlab{a}}, MNRAS,
  413, 1165

\bibitem[{Curran \& Whiting(2010)}]{cw10}
Curran, S.~J. \& Whiting, M.~T. 2010, ApJ, 712, 303

\bibitem[{Curran {et~al.}(2011{\natexlab{b}})Curran, Whiting, Tanna, Bignell,
  \& Webb}]{cwt+11}
Curran, S.~J., Whiting, M.~T., Tanna, A., Bignell, C., \& Webb, J.~K.
  2011{\natexlab{b}}, MNRAS, 413, L86

\bibitem[{Curran {et~al.}(2011{\natexlab{c}})Curran, Whiting, Webb, \&
  Athreya}]{cwwa11}
Curran, S.~J., Whiting, M.~T., Webb, J.~K., \& Athreya, A. 2011{\natexlab{c}},
  MNRAS, 414, L26

\bibitem[{Curran {et~al.}(2008)Curran, Whiting, Wiklind, Webb, Murphy, \&
  Purcell}]{cww+08}
Curran, S.~J., Whiting, M.~T., Wiklind, T., Webb, J.~K., Murphy, M.~T., \&
  Purcell, C.~R. 2008, MNRAS, 391, 765

\bibitem[{Hogg(1999)}]{hog00}
Hogg, D.~W. 1999, astro-ph/9905116

\bibitem[{{Isobe} {et~al.}(1986){Isobe}, {Feigelson}, \& {Nelson}}]{ifn86}
{Isobe}, T., {Feigelson}, E., \& {Nelson}, P. 1986, ApJ, 306, 490

\bibitem[{Kanekar \& Chengalur(2001)}]{kc01a}
Kanekar, N. \& Chengalur, J.~N. 2001, A\&A, 369, 42

\bibitem[{Kanekar \& Chengalur(2003)}]{kc02}
---. 2003, A\&A, 399, 857

\bibitem[{{Kanekar} {et~al.}(2009){Kanekar}, {Lane}, {Momjian}, {Briggs}, \&
  {Chengalur}}]{klm+09}
{Kanekar}, N., {Lane}, W.~M., {Momjian}, E., {Briggs}, F.~H., \& {Chengalur},
  J.~N. 2009, MNRAS, 394, L61

\bibitem[{{Lavalley} {et~al.}(1992){Lavalley}, {Isobe}, \&
  {Feigelson}}]{lif92a}
{Lavalley}, M.~P., {Isobe}, T., \& {Feigelson}, E.~D. 1992, in BAAS, Vol.~24,
  839--840

\bibitem[{Murphy {et~al.}(2007)Murphy, Curran, Webb, M\'{e}nager, \&
  Zych}]{mcw+07}
Murphy, M.~T., Curran, S.~J., Webb, J.~K., M\'{e}nager, H., \& Zych, B.~J.
  2007, MNRAS, 376, 673

\bibitem[{{Noterdaeme} {et~al.}(2009){Noterdaeme}, {Petitjean}, {Ledoux}, \&
  {Srianand}}]{npls09}
{Noterdaeme}, P., {Petitjean}, P., {Ledoux}, C., \& {Srianand}, R. 2009, A\&A,
  505, 1087

\bibitem[{Prochaska {et~al.}(2003)Prochaska, Gawiser, Wolfe, Castro, \&
  Djorgovski}]{pgw+03}
Prochaska, J.~X., Gawiser, E., Wolfe, A.~M., Castro, S., \& Djorgovski, S.~G.
  2003, ApJ, 595, L9

\bibitem[{Prochaska {et~al.}(2005)Prochaska, Herbert-Fort, \& Wolfe}]{phw05}
Prochaska, J.~X., Herbert-Fort, S., \& Wolfe, A.~M. 2005, ApJ, 635, 123

\bibitem[{{Roberts} {et~al.}(1976){Roberts}, {Brown}, {Brundage}, {Rots},
  {Haynes}, \& {Wolfe}}]{rbb+76}
{Roberts}, M.~S., {Brown}, R.~L., {Brundage}, W.~D., {Rots}, A.~H., {Haynes},
  M.~P., \& {Wolfe}, A.~M. 1976, AJ, 81, 293

\bibitem[{{Srianand} {et~al.}(2012){Srianand}, {Gupta}, {Petitjean},
  {Noterdaeme}, {Ledoux}, {Salter}, \& {Saikia}}]{sgp+12}
{Srianand}, R., {Gupta}, N., {Petitjean}, P., {Noterdaeme}, P., {Ledoux}, C.,
  {Salter}, C.~J., \& {Saikia}, D.~J. 2012, MNRAS, accepted (arXiv:1112.1438)

\bibitem[{{Wolfe} \& {Burbidge}(1975)}]{wb75}
{Wolfe}, A.~M. \& {Burbidge}, G.~R. 1975, ApJ, 200, 548

\bibitem[{{Wolfe} {et~al.}(2005){Wolfe}, {Gawiser}, \& {Prochaska}}]{wgp05}
{Wolfe}, A.~M., {Gawiser}, E., \& {Prochaska}, J.~X. 2005, ARA\&A, 43, 861

\end{thebibliography}

\end{document}